Deepfakes and the 2020 US elections: what (did not) happen

João Paulo Meneses, CECS, Portugal

D011454@ismai.pt

Abstract

Alarmed by the volume of disinformation that was assumed to have taken place during the 2016 US elections, scholars, politics and journalists predicted the worst when the first deepfakes began to emerge in 2018. After all, US Elections 2020 were believed to be the most secure in American history. This paper seeks explanations for an apparent contradiction: we believe that it was precisely the multiplication and conjugation of different types of warnings and fears that created the conditions that prevented malicious political deepfakes from affecting the 2020 US elections. From these warnings, we identified four factors (more active role of social networks, new laws, difficulties in accessing Artificial Intelligence and better awareness of society). But while this formula has proven to be effective in the case of the United States, 2020, it is not correct to assume that it can be repeated in other political contexts.

João Paulo Meneses

C..EC.S., Portugal

In

E-mail: d011454@ismai.pt

ORCID: 0000-0003-2365-3832

**Introduction**

One year before the US Elections 20, Nisos[1] published a report ('Understanding the illicit economy for synthetic media'[2]) that read 'we do not anticipate widespread deepfake[3] use in disinformation campaigns in the near term (to include the 2020 election cycle)'.

In retrospect, Nisos experts made the right forecast. However, this was a clear minority opinion. Before and after their report, dozens of politicians and institutions drew considerable attention to the approaching danger: 'imagine a scenario where, on the eve of next year's presidential election, the Democratic nominee appears in a video where he or she endorses President Trump. Now, imagine it the other way around.' (Sprangler, 2019).

It is fair to say that deepfakes' high potential for disinformation was noticed long before these hypothetical consequences were evoked, mainly because they were revealed to be highly credible. Two examples: 'In an online quiz, 49 percent of people who visited our site said they incorrectly believed Nixon's synthetically altered face was real and 65 percent thought his voice was real' (Panetta et al, 2020), or 'Two-thirds of participants believed that one day it would be impossible to discern a real video from a fake one. 42 percent of people believed it is very or extremely likely that deepfakes will be used to mislead voters in 2020' (AMEinfo, 2020).

However, perhaps the most frightening factor to anyone interested in the consequences of disinformation was the predictable difficulty in combating or neutralizing malicious deepfakes, since they are built on a technology that is capable of adapting and evolving: Artificial Intelligence (AI).

'Expert says detecting deepfakes almost impossible', was the suggestive headline of an Axios story (Fernandez, 2020). The same expert, Hany Farid, a professor from University of California, Berkeley, stated in other source that '[i]n January 2019, deepfakes were buggy and flickery. Nine months later, I've never seen anything like how fast they're going. This is the tip of the iceberg' (Toews, 2020).

Many factors make this technology something special and probably nothing compared to what was known until now, as Katarya and Lal (2020) state: 'The existence of such open-source software and the availability of devices in the market for fabricating and propagating these falsified information has brought to attention the immediate need for detection and elimination of malicious deepfake content'.

### 1.1 Warnings about the 2020 US elections

The dangers that deepfakes can pose for an election were widely anticipated prior to the 2020 US elections. Although the weight of the two technologies is different, the basic fear was that deepfakes could play the same role that fake news did in 2016.[4] Consequently, there were warnings from various sectors (mainly politicians, technologists, and academics) about the potential of deepfakes to cause irreparable harm. Much more because 'with so few people undecided about the upcoming presidential election, influencing just a handful of people on the margins can sway an election' (Polakovic, 2020). The following list presents some relevant examples of this potential:

- 'In the old days, if you wanted to threaten the United States, you needed 10 aircraft carriers, and nuclear weapons, and long-range missiles. Today (…) all you need is the ability to produce a very realistic fake video that could undermine our elections, that could throw our country into tremendous crisis internally and weaken us deeply,'stated US Senator Marco Rubio (Porup, 2019).
- 'In his opening remarks, committee Chair Adam Schiff warned of a 'nightmarish' scenario for the upcoming presidential campaigns and declared that 'now is the time for social media companies to put in place policies to protect users from misinformation, not in 2021 after viral deepfakes have polluted the 2020 elections'' (Galtson, 2020).
- Senator Ben Sasse, a Republican from Nebraska who introduced a bill to criminalize the malicious creation of deepfakes, warned in 2019 that the technology could 'destroy human lives', 'roil financial markets', and even 'spur military conflicts around the world' (Wolfgang, 2018).
- 'How exactly could deepfakes be weaponized in an election? To begin with, malicious actors could forge evidence to fuel false accusation and fake narratives. (…) Deepfakes could also be used to create entirely new fictitious content, including controversial or hateful statements with the intention of playing upon political divisions, or even inciting violence', wrote researchers Puutio and Timis (2020).
- 'Disinformation conveyed via deepfakes could pose a challenge during elections, since, to the untrained eye, a deepfake may be difficult to distinguish from a real video. Any political actor could try to discredit an opponent or try to incite some political scandal with the goal of furthering their own agenda' (Dobber et al, 2020: 2).
- 'The upcoming US presidential election in November 2020 will serve as a bellwether not only for Western liberal democracies, but for the rest of the world' (Schick, 2020a: 21) and 'I believe the corrupt information ecosystem will play an even greater role in the 2020 election than it did in 2016' (Schick, 2020a: 113).
- 'Deepfake videos and audios could undermine the democratic process by tipping an election', said Danielle Citron, a then-law professor at the University of Maryland (Simonite, 2020).
- 'If executed and timed well enough, such interventions are bound to tip an outcome sooner or later—and in a larger set of cases they will at least cast a shadow of illegitimacy over the election process itself' (Citron & Chesnet, 2019, 1779).
- ''Democracies will be at a disadvantage' if deepfakes become common, Emerging Technologies Fellow Lindsay Gorman told panelists gathered online, on March 12 at the Information Technology Innovation Foundation' (Patton, 2020).
- 'We've seen proofs of concepts of deepfakes being released that could be used to influence the electorate. (…) If a deepfake is dropped at the right time, say maybe two or three days before an election occurs, imagine the impact that could have if it goes viral?', stated Matt Price of ZeroFox (Roby, 2019).
- 'Artificial intelligence is the real thing. It is already in use by attackers. When they learn how to do deepfakes, I would argue this is potentially an existential threat', said Major-General (ret.) Brett Williams, the former US cyber command chief (Grossman, 2020).

---

[1] https://www.nisos.com/company
[2] https://www.nisos.com/deep_fakes
[3] In our definition, a deepfake is completely or partially fake content in video, audio, text and/or image form that was generated using Artificial Intelligence. Thus, deepfakes are not limited to their most popular form, namely videos
[4] In summary, the Mueller Report concluded that Russia's Internet Research Agency and military intelligence service (GRU) used a range of digital tactics to target the 2016 US elections.

This list of examples is intended to show that public warnings about the dangers of deepfakes have appeared since at least 2018. On the other hand, the issue united people with very different interests around a common goal.

**1.2 Some previous cases**

'We haven't seen any deepfakes released in the wild that we think are genuinely malicious, not saying that they're deepfakes and trying to mask what they are', said Price (Roby, 2019). In addition, the cases that went public in the two years prior to the 2020 elections were isolated or even anecdotal.

In January 2019 in Gabon, a video of President Ali Bongo, who had not made a public appearance for several months, triggered a coup. 'The military believed the video was a fake, although the president later confirmed it was real' (Siyech, 2020).

In India, a day before the Delhi election in February, two videos of Delhi unit Bharatiya Janata Party (BJP) President Manoj Tiwari hit the internet where he was found urging voters to vote to his party. The videos were then reported as deepfakes (Kumar, 2020)

In Italy, an Italian satirical television show used a deepfake video unfavourable to the prime minister, Matteo Renzi. 'The video shared on social media depicted him insulting fellow politicians. As the video spread online, many individuals started to believe the video was authentic, which led to public outrage' (Buo, 2020).

However, there appears to be a disconnect between this succession of warnings and the events that subsequently transpired. As Grossman (2020) expressed, 'if there has been a surprise in campaign tactics this cycle, it is that these AI-generated videos have played a very minor role, little more than a cameo'. This paper intends to explain this gap by not only seeking answers but also suggesting recommendations, using historical and secondary research.

**2.0 What really happened in US Election 20**

Collectively, there are four reasons that explain why malicious political deepfakes did not appear during the electoral period in the United States. Over the course of at least during the year before the elections, thousands of deepfakes were disseminated; although most were of a pornographic nature, hundreds of videos that were seemingly benign in scope (e.g., humorous, artistic, etc.) were also released. In other words, not only did deepfake technology continue to develop, but hundreds of creators[5] around the world continued to produce this type of content.

**2.1 Interventions from social networks and technological platforms**

After receiving significant criticism for their passivity in the 2016 US elections, social networks (and technological platforms more generally) changed their discourse from 2018 onwards. They moved away from maintaining a generally neutral attitude towards the content and towards a more active, interventionist, and even controversial form of censorship.

> 'If they do not avail themselves of this opportunity—and if deepfakes rampage through next year's election, leaving a swathe of falsehoods and misrepresentations in their wake—Congress may well move to strip the platforms of the near total immunity they have enjoyed for a quarter of a century, and the courts may rethink interpretations of the First Amendment that prevent lawmakers from protecting fundamental democratic processes' (Chesney and Citron, apud Galston, 2020).

---

[5] https://www.technologyreview.com/2019/09/25/132884/google-has-released-a-giant-database-of-deepfakes-to-help-fight-deepfakes/

Hwang (2020: 14) states that 'Public and policymaker concern has led to many of these platforms declaring policies against uploading deepfake content under certain circumstances. Furthermore, 'there are no laws that regulate what's going on with social media and websites today. (…) they don't really have an incentive to go around and try to take down this synthetic content or even note that it is synthetic content' (Roby, 2019). And the chief executive officer of Alphabet and Google, Sundar Pichai (2020), underlines: 'Now there is no question in my mind that artificial intelligence needs to be regulated. It is too important not to. The only question is how to approach it'.

Some of the most significant moves of social networks were as follows:

- In January 2020, Facebook announced that they would begin removing content that 'has been edited or synthesized (…) in ways that aren't apparent to an average person and would likely mislead someone into thinking that a subject of the video said words that they did not actually say' (Hwang, 2020: 14); 'Facebook claims A.I. Now Detects 94.7% of the Hate Speech That Gets Removed From Its Platform' (Shead, 2020).
- Twitter adopted a broader approach in February 2020, announcing that they would remove and warn users against 'synthetic or manipulated media that are likely to cause harm' (Hwang, 2020: 14). A Twitter blog post last week rounding up its election efforts said 'it had added labels warning of misleading content to 300,000 tweets since October 27, which was 0.2 percent of all election-related posts in that period. It didn't mention deepfakes' (Simonite, 2020).
- Platforms such as Reddit have adopted similar policies (Hwang, 2020: 14).
- In September 2019, Facebook announced a $10 million 'Deepfake Detection Challenge'. This was a partnership with six leading academic institutions and companies, including Amazon and Microsoft (Galston, 2020). 'These efforts will likely play a significant role both in pushing research forward and smoothing the progression of technical advancements into practical software solutions that can be implemented by intermediaries across the web' (Hwang, 2020: 15).

The above list shows that the main social networks chose to take a more active role in the fight against deepfakes. However, others did not, which led Camille François to remark that

'while Facebook, Twitter, and Google have drawn intense scrutiny, many platforms who consider themselves outside of the political conversation and other alternative platforms designed to host content moderated away from the main platforms carry the risk of creating an entire alternative ecosystem where disinformation and hate can thrive' (TSPAF, 2020).

### 2.1.1 Seeking technological solutions to combat deepfakes

To combat fake news, social networks resorted to fact-checking. However, eliminating malicious deepfakes requires – and will increasingly require – the use of AI tools. Research by Ahmed (2020), Farid and Agarwal (Manke, 2019), and Thaw et al. (2020) demonstrated that ordinary people had difficulty identifying videos made with deepfake technology.

There have been countless efforts to develop effective detection systems (Goled, 2020), but 'have yet to establish a foolproof method' (Grossmann, 2020). 'Factors such as the need to avoid attribution, the time needed to train a Machine Learning model, and the availability of data will constrain how sophisticated actors use tailored deepfakes in practice', suggested Hwang (2020: iii). Furthermore, Du et al. (2020) recognized that 'although lots of efforts have been devoted to detect deepfakes, their performance drops significantly on previously unseen but related manipulations and the detection generalization capability remains a problem'.

> 'It's a good thing that the 2020 election wasn't swarmed by deepfakes, because attempts to automatically detect AI-generated video haven't yet been very successful. When Facebook challenged researchers to create code to spot deepfakes, the winner, announced in June, missed more than a third of deepfakes in its test collection' Simonite (2020).

### 2.2 New laws about deepfakes

Once again, the lesson of 2016: before deepfakes could be a problem in the 2020 US elections, laws were created to stop them. Efforts were made to ensure that the laws immediately had a federal dimension, but concrete results actually manifested in two states: California and Texas.

California's case is more relevant, not only due to its stature as the largest state in the United States (and home to many technology companies), but also because two laws were passed that came into effect in early 2020. The first law (AB 602) was designed to combat pornographic deepfakes, whilst the second law (AB 730) prohibited the use of deepfakes to influence political campaigns. However, AB730 will lapse on 1 January 2023.

AB 730 prohibits the distribution of materially deceptive audio or visual media that depicts a candidate for office within 60 days of an election 'with actual malice', or the intent to injure the candidate's reputation or deceive voters into voting for or against the candidate. 'Significantly, this measure exempts print and online media and websites if that entity clearly discloses that the deepfake video or audio file is inaccurate or of questionable authenticity' (Halm et al, 2019).

Along the way, efforts were made to to create legislation by lawmakers such as Ben Sasse, who introduced what was viewed as the first bill to criminalize the malicious creation and distribution of deepfakes. Introduced a day before the government shutdown, the bill flew under the radar and expired when the year ended (Waddell, 2019). In addition, Representative Yvette Clark introduced the DEEPFAKES Accountability Act in June 2019. 'This bill would make it a crime to create and distribute a deepfake without including a digital marker of the modification and text statement acknowledging the modification' (Galston, 2020).

On 20 December 2019, President Donald Trump signed the United States' first federal law related to deepfakes. This legislation was part of the National Defense Authorization Act (NDAA) for Fiscal Year 2020. In two provisions related to this emerging technology, the NDAA (1) requires a comprehensive report on the foreign weaponization of deepfakes; (2) requires the government to notify Congress of foreign deepfake-disinformation activities targeting US elections; and (3) establishes a 'Deepfakes Prize' competition to encourage the research or commercialization of deepfake-detection technologies,' as explained by Haleet al (2019). These measures resulted in the Deepfakes Report Act of 2019.[6]

Although it is not legislation, we can associate the webpage 'Rumor vs. Reality'[7] sponsored by the Cybersecurity and Infrastructure Security Agency to this topic. The agency is part of the Department of Homeland Security, which continued to be active after November 3. Or the two programs created by the Defense Advanced Research Projects Agency, intended to improve defenses against adversary information operations', states the Congressional Research Service (Neus, 2020).

---

[6] https://www.govtrack.us/congress/bills/116/hr3600
[7] https://www.cisa.gov/rumorcontrol

### 2.3 'Why take the risk?'

Of the four reasons presented in this paper, there is one that, at least in part, does not result from such a concerted but unorganized effort to prevent deepfakes: 'depending on what the deepfake depicts, there may be significant expense in acquiring the training data, structuring it properly, and running the training process' (Hwang, 2020: 3).

Whilst some who bet on disinformation as a tactic in electoral interference were intimidated by these efforts (either because they feared the consequences Of being caught or that their work would be halted at the source, namely on social networks), others understood that resorting to deepfake technology was not necessary or even possible.

'Those who would use deepfakes as part of an online attack have not yet mastered the technology, or at least not how to avoid any breadcrumbs that would lead back to the perpetrator', wrote Gary Grossman (2020). However, 'deepfakes have earned a reputation for unique and deadly precision' (Simonite, 2020), most still aren't that good. 'For a while yet it is likely to be easier to fake reality using conventional video production techniques' (Simonite, 2020). And if deepfakes haven't arrived until US Elections 20, 'it's not because they aren't a threat (…). It's because simple deceptions like selective editing or outright lies have worked just fine' (Mak & Temple-Raston, 2020).

In other words, because 'factors such as the need to avoid attribution, the time needed to train a Machine Learning model, and the availability of data will constrain how sophisticated actors use tailored deepfakes in practice' (Hwang, 2020: iv) or just because 'people seem to be much more taken with pornographic possibilities than bringing down Governments' (Boyd, 2020), the truth is that the idea that '[u]ploading an algorithmically doctored video is likely to attract attention from automated filters, while conventional film editing and obvious lies won't. Why take the risk?' (Brandom, 2019).

### 2.4 Greater social awareness of disinformation

Although this may be the least relevant factor, it may be appropriate to highlight what appears to be a greater social awareness of disinformation, especially compared to the period before the 2016 US elections. This change began with universities, which made the topic of disinformation a priority (Katarya et al, 2020). In parallel, 'there is a rising number of newly formed startups claiming to be able to tackle the problem of deepfakes' (Stacey, 2019).

The efforts of the Knight Foundation, a freedom of speech advocate, are also worth noting. The organization awarded $50 million in grants in 2019 to 11 universities and research institutions in the United States to study how technology is transforming democracy (Brunner, 2019).

One of the curious observations about deepfakes affirmation process is that, at different times and in different countries, deepfake technology has been used to denounce the dangers of deepfake technology itself. There are many examples of this, including the classic deepfake produced in 2018 by comedian Jordan Peele, in which former US President Barack Obama calls Trump as a 'dipshit'[8]; the two videos commissioned by British think tank Future Advocacy, in which British Prime Minister Boris Johnson endorses his opponent, Jeremy Corbyn; and another video in which Corbyn endorses Johnson[9]; or a pair of deepfake advertisements – perhaps the first serious use of deepfakes in the US political campaign – released by nonpartisan advocacy group RepresentUs featuring Russian President Vladimir Putin and North Korean leader Kim Jong-un, both disseminating the same message: they do not need to interfere with the US elections, because the United States was capable of damaging its democracy on its own.[10] *'Some have even been used as part of a public service campaign to express the importance of saving democracy'* (Grossmann, 2020).

---
[8] https://www.youtube.com/watch?v=bE1KWpoX9Hk

Lastly, *'the fact-checking and journalism ecosystem did better with 2020 disinformation than many had feared after the distortions of 2016'* (Simonite, 2020) (Chaturvedi, 2020) or (Spencer, 2020).

**3.0 Conclusion**

*'The November 3rd election was the most secure in American history'*, claimed the Election Infrastructure Government Coordinating Council executive committee and the Election Infrastructure Sector Coordinating Council[11].

The question of this paper is not about disinformation in general, since fake news continued to proliferate in the months prior to the elections (Abbasi and Derakhti, 2020; Schick, 2020a: 75). Rather, it is about malicious political deepfakes. They appear to have played a minor role in US Elections 2020, which was indicated by the fact that the most-discussed topic in the months before 3 November was a faked video showing former Vice President Joe Biden sticking his tongue out, which was tweeted out by the president himsel (Mak and Temple-Raston, 2020). A more relevant case, on several levels, was a dossier at Typhoon Investigations detailing the hypothetical Chinese business connections of Hunter Biden, Joe Biden's son. Eventually, journalists denounced the report as being authored by a fake person, *'the spurious photograph of whom is a realistic avatar created by AI. The 64-page 'Martin Aspen' dossier is a fraudulent project, fraught with slimy intrigue throughout'* (Cunningham, 2020)

From our point of view, the previously designed scenario created an environment of hypersensitivity, which led to social networks to not only be much more attentive to desinformations and proliferation of deepfakes but also to intervene, preventing it – a measure that they had not taken until the moment when the elections were approaching – and to the creation of laws to prevent what really had not yet happened.

Can we consider warnings made in 2019 and 2020 regarding the use of deepfakes in US elections to be exaggerated or even premature? Apparently so, but it was this exaggeration that enabled the creation of a climate that helped to nullify the putative effect of deepfakes.

It was a combination of various factors, as if we faced a (positive) *'perfect storm'*, that made the 2020 elections in the United States the *'most secure'*. Some of these factors may have been more relevant than others, as noted by Chesney and Citron, who believed that the content screening and removal policies of social media platforms may have proven to be *'the most salient response mechanism of all'*, as their terms of service agreements are *'the single most important documents governing digital speech in today's world'* (Galston, 2020).

---

[9] https://futureadvocacy.com/deepfakes/
[10] https://act.represent.us/sign/deepfake-release/
[11] https://www.cisa.gov/news/2020/11/12/joint-statement-elections-infrastructure-government-coordinating-council-election

However, we followed Paul Barrett, the author of a New York University 2019 report that listed deepfakes first on a list of disinformation predictions for 2020.[12] He wrote that *'the warnings may have worked, convincing would-be deepfake producers that their clips would be quickly unmasked. But he warns that the threat remains'* (Simonite, 2020). In addition, Paris and Donovan argued that an exclusively technological approach is insufficient for addressing the threats of deep and cheap fakes[13]. According to them, any solution must take into account both *'the history of evidence'* (2019:44) and the *'social processes that produce truth'* (2019:47). Furthermore, Price stated that *'there's a number of different ways that this problem can be tackled, and I don't think anyone by itself is a solution'* (Roby, 2019).

We agree with Grossmann (2020) when he states, *'the reason there have not been more politically motivated malevolent deepfakes designed to stoke oppression, division, and violence is a matter of conjecture'*. The abovementioned combination of factors resulted in this specific context, but there is no guarantee that the same will be true of the future, in other elections. The public and widespread perception in the United States of the role that disinformation played in the 2016 elections was very important. In another electoral context, this widespread perception is unlikely to exist, since *'across most of the world there is minimal commitment to taking on malicious deepfakes'* (Lamb, 2020).

### 3.1 Recommendations

Given the challenging nature of policing deepfakes and the fact that they are *'also the most sophisticated tool of misinformation and disinformation that has existed to date'* (Carruthers, 2020), the following list of recommendations is aimed at helping to address this problem in the future:

- Continuing to develop legislation aimed at stopping the creation of malicious deepfakes without affecting freedom of expression. *'Deepfakes have indeed blurred the lines of truth and pushed us towards a post-truth world'* (Raj, 2020).
- Using available systems in the absence of a universal detection and blocking system. This would allow to improve the final quality of the best detection software. *'As the deepfake technology approaches towards generating fake content with considerably improved quality, it will likely become impossible to detect them shortly'* (Katarya and Lal, 2020).
- Continuing to invest in increasingly effective detection and blocking systems. This will also depend on the level of funding that universities and startups can secure. As stated by Nasir Memon, a professor of computer science at NYU, *'there's no money to be made out of detecting [fake media], which means that few people are motivated to research ways to detect deepfakes'* (Redick, 2020).
- *'Also of particular concern is the use of deep fakes in propaganda and misinformation in regions with fragile governance and underlying ethnic tensions. (…) This highly divisive content spreads quickly because it appeals to emotions'* (Smith and Mansted, 2020).

We must be *'always be skeptical, regardless of suspicions over AI involvement. The truth is most definitely out there… it just might take a little longer to reach than usual'*, stated Boyd (2020). This is more important because *'new security measures consistently catch many deepfake images and videos, people may be lulled into a false sense of security and believe we have the situation under control', said Professor Bart Kosko of the Ming Hsieh Department of Electrical and Computer Engineering; University of South California (Paul, 2020). This must avoid at all costs. In addition, as Schick (2020) anticipated, *'the question, then, should not be '(When) will political deepfakes emerge?' but 'How we can mitigate the many ways in which visual disinformation is already reshaping our political reality?'*

---

[12] https://bhr.stern.nyu.edu/tech-disinfo-and-2020-election
[13] Videos manipulated without resorting to Artificial Intelligence

### 3.2 Limitations

The writing of this paper began shortly after 3 November, the US Election day, and concluded less than three months later. Naturally, more and better information about the situation is bound to emerge at a later date.

Although it was stated in this paper that there was no knowledge of malicious political deepfakes during the two years prior to the 2020 elections in the United States, it is important to note that several deepfakes were very likely stopped by social network detection algorithms before they could be released and shared – that is, they would have existed if not for these systems, but they were pre-emptively identified and the public never learned of their existence.